\documentclass{article}

\usepackage{soul}
\usepackage{color}
\usepackage{lineno}
\usepackage{wrapfig}
\usepackage[authoryear]{natbib} 
\usepackage[margin = 1in]{geometry}
\usepackage{authblk}
\usepackage{graphicx}
\usepackage{textcomp}

\begin{document}

% Main title of the paper
\title{Contemporary climate analogs project north–south polarization of urban water-energy nexus across US cities under warming climate}  

\author[1,2,3,*]{Renee Obringer}
\author[4]{Roshanak Nateghi}
\author[4]{Jessica Knee}
\author[5,6]{Kaveh Madani} 
\author[7,*]{Rohini Kumar} 

% Renee:
\affil[1]{Department of Energy and Mineral Engineering, Pennsylvania State University, University Park, PA 16802, United States}

\affil[2]{National Socio-Environmental Synthesis Center, University of Maryland, Annapolis, MD 21401, United States}

\affil[3]{Ecological Science and Engineering, Purdue University, West Lafayette, IN 47907, United States}

% Roshi/Jessica:
\affil[4]{School of Industrial Engineering, Purdue University, West Lafayette, IN 47907 United States}

% Kaveh:
\affil[5]{United Nations University Institute for Water, Environment and Health (UNU-INWEH), Hamilton, ON Ontario L8P 0A1, Canada}

\affil[6]{The City University of New York Remote Sensing of Earth Science and Technology (CUNY-CREST) Institute, New York City, NY 10031, United States}

% Rohini:
\affil[7]{Department of Computational Hydrosystems, Helmholtz Centre for Environmental Research - UFZ, Leipzig 04318, Germany}

\affil[*]{Corresponding authors: obringer@psu.edu; rohini.kumar@ufz.de}

\date{}

\maketitle

\begin{abstract}
Despite the coupled nature of water and electricity demand, the two utilities are often managed by different entities with minimal interaction. Neglecting the water-energy demand nexus leads to to suboptimal management decisions, particularly under climate change. Here, we leverage state-of-the-art machine learning and contemporary climate analogs to project the city-level coupled water and electricity demand of 46 major U.S. cities into the future. The results show that many U.S. cities may experience an increase in electricity (water) demand of up to 20\% (15\%) due to climate change under a high emissions scenario, with a clear north-south gradient. In the absence of appropriate mitigation strategies, these changes will likely stress current infrastructure, limiting the effectiveness of the ongoing grid decarbonization efforts. In the event that cities are unable to match the increasing demand, there may be increased occurrence of supply shortages, leading to blackouts with disproportionate impacts on vulnerable populations. As such, reliable projections of future water and electricity demand under climate change are critical not only for preventing further exacerbation of the existing environmental injustices but also for more effective design and execution of climate change mitigation and adaptation plans.
\end{abstract}

\section*{Introduction} 
Globally, cities are growing at an increasingly fast rate, with nearly 70\% of the world's population estimated to live in cities by 2050 \citep{unitednations2018}. As urban populations  grow, so will the need for critical services such as water and electricity. In fact, the provisioning of these services is of paramount importance to sustainable development \citep{unitednations2015}. The compound effects of population growth and anthropogenic climate change will constrain the supply of water and electricity and yet increase the demand for these resources, straining the existing water and energy infrastructure systems \citep{connell-buck2011, cronin2018, auffhammer2017, raymond2019, ashraf2019, mukherjee2019a, obringer2020a}.  For this reason, there is a major area of research that focuses on the impact of climate change on the water-energy nexus. These studies primarily focus on the supply-side, such as the nexus between water supply and electricity generation through hydropower \citep{guegan2012, madani2014} or the use of water by electricity generators for cooling purposes \citep{macknick2012a,dale2015}. There is, however, a growing body of work that investigates the demand-side of the nexus, looking in particular at the simultaneous consumption of water and electricity \citep{bartos2014c,escriva-bou2018,obringer2019c}. 

Within the literature on the urban water-energy nexus \citep{newell2019, newell2020}, the focus tends to be on smaller scales, such as a single city \citep{dale2015,escriva-bou2018} or a small group of cities \citep{venkatesh2014a,obringer2019c,obringer2020a}. Given that management of water and electricity systems occurs at multiple scales (e.g., ranging from local utilities to federal management entities), there is a need to expand urban water-energy nexus analyses to consider larger scales. 

Additionally, studies that leverage data-driven modeling of climate change impacts on the \textit{climate-sensitive portion of} end-use demand primarily involve first characterizing the relationship between climate and demand in the observational space, then projecting those relationships into the future using General Circulation Model- (GCM-) derived climate data \citep{mukherjee2017a, mukherjee2019, maia-silva2020, obringer2020a, obringer2022}. While this practice is beneficial and can provide accurate representations of future demand, the incorporation of GCM-based data is computationally expensive, rendering it unsuitable for local utilities without access to high-performance computing and necessary expertise. 

Overall, it is critical that urban water-energy nexus studies account for anthropogenic climate change, which is one of the most pressing challenges society faces. To do this, there is a need to create transferable frameworks that can be readily applied to large-scale analyses. Additionally, it is beneficial, particularly from a practitioner's perspective, to evaluate quick, but reliable techniques for obtaining future climate data that are more user-friendly and less computationally-intensive, such as the climate analogs approach.

Climate analogs are determined through statistical processes that identify the location in which the \textit{current} climate is analogous to the \textit{future} climate in another location. In particular, the climate analogs used in this study, which were developed by \citet{fitzpatrick2019}, used 27 GCMs to evaluate the climate analogs for 540 cities in North America, based on both temperature and precipitation. In particular, the original study leveraged minimum and maximum temperature and total precipitation for the four climatological seasons to conduct their analysis \citep{fitzpatrick2019}. These analogs were calculated based on the sigma dissimilarity between these twelve variables (3 measures $\times$ 4 seasons) in the present-day city's and the future city's climate. Thus, each present-day city has 27 possible analogs using each of the selected GCMs, plus an ensemble analog. In this study, we used the ensemble analogs for 46 major US cities.

The analog approach has since been adopted in other locations, such as West Africa \citep{fitzpatrick2020} and China \citep{yin2020, wang2022}, as well as for communication and policy studies \citep{luccioni2021, binelli2022}. However, it should be noted that, as with any approach that leverages future climate simulations, there is uncertainty associated with the climate analogs, as there is no `perfect match'. In fact, \citet{fitzpatrick2019} found that in some cities, the average analog was in a completely different climatic zone, indicating that extreme climate change could lead to unexpected changes. 

In this work, we leverage climate analogs to characterize the future climate in 46 major cities (population $>$ 250,000) across the United States. These cities were chosen based on data availability out of an initial pool of 80 cities that met our population criterion. Using a state-of-the-art machine learning model (i.e., multivariate tree boosting \citep{miller2016}), we use the observed climate data in the analogs to project the coupled water and electricity demand for each city of interest. The goal of this study is to focus on the impact of climate change on urban utilities. In particular, we present the `worst case scenario', in which there is no adaptation within the city to account for the changing climate conditions and no mitigation to reduce carbon emissions at the global level. Additionally, we present results using the high emissions (RCP8.5) climate analogs. See Supplementary Figure S2 for the results from a moderate emissions scenario  (RCP4.5).

This study aims to fill a significant gap in the literature by presenting a computationally inexpensive projection of the climate-related changes in future water and electricity demand across a number of major U.S. cities, while also emphasizing the importance of considering the water-electricity nexus in such analyses. Additionally, by using climate analogs as the primary source of future climate data we aim to communicate climate change impacts to US citizens and decision-makers in a more intuitive way than traditional GCM-based studies that often do not have a comparative analysis basis. 

In the following sections we will discuss the results of using climate analogs to project water and electricity demand across the United States. We will first focus on the large-scale regional trends before delving into specific cases of interest, such as exploring the north-south gradient and evaluating the impact of various shared socioeconomic pathways on the total water and electricity demand. We will wrap up with a brief discussion of the implications of the results and the limitations of this study.

\section*{Methods}

\subsection*{Data}
In this study there were two types of data: consumption data and climate data. The consumption data included both water and electricity consumption, which were obtained from different sources. The water consumption data were collected directly from the water utilities in the selected cities via Freedom of Information Act (FOIA) requests. Initially, cities with populations greater than 250,000 were selected, resulting in 80 initial cities. Of these 80 cities, only 46 cities (see Supplementary Table S1 or Figure~\ref{fig2}a) responded with adequate data for our analysis. In particular, some cities simply did not respond the the request, while others were unable to provide monthly water consumption for the time period of interest---2007 to 2018. The electricity consumption data were obtained from the US Energy Information Administration (EIA), which keeps values of monthly electricity use for a number of utilities across the US \citep{usenergyinformationadministration2019}. It should be noted that in the case of Texas, many of the utilities were not included in the EIA dataset, which led to some Texas cities being removed from the analysis. Finally, both water consumption and electricity consumption were normalized by service population and de-trended to remove the impact non-climatic factors (i.e., technology, socioeconomic status) following a method by Sailor and Mu\~{n}oz \citep{sailor1997}. The normalization process was important to allow for a closer comparison between cities, as well as the water and electricity consumption within a city, as many water and electric utilities have different jurisdictions. The de-trending process, allowed us to focus on the climate-sensitive portion of the water and electricity consumption, rather than including the technological and socioeconomic factors that also influence the demand profiles. 

The climate data were collected from the North American Regional Reanalysis \citep{mesinger2006}. The variables included dry bulb temperature, wet bulb temperature, dew point temperature, relative humidity, wind speed, and precipitation. Initially, the data were collected as daily values for the entire time period in each city, but were later aggregated to monthly variables. In particular, the temperature, humidity, and wind speed variables were aggregated to form minimum, maximum, and mean monthly values, while the precipitation as aggregated to be the accumulated monthly value, as well as the number of days within the month in which there was precipitation. Previous work has leveraged the NARR dataset for observational climate data, following a similar collection process as this study \citep{maia-silva2020}.  Since we used the climate analog approach, there was no need to run any simulations to obtain the future climate variables. Rather, the climate variables from the various analog cities were used to represent the future climate of the cities of interest. See the Supplementary Information for more information on the development of the climate analogs.

\begin{figure}[t!]
\centering
\includegraphics[width=.73\linewidth]{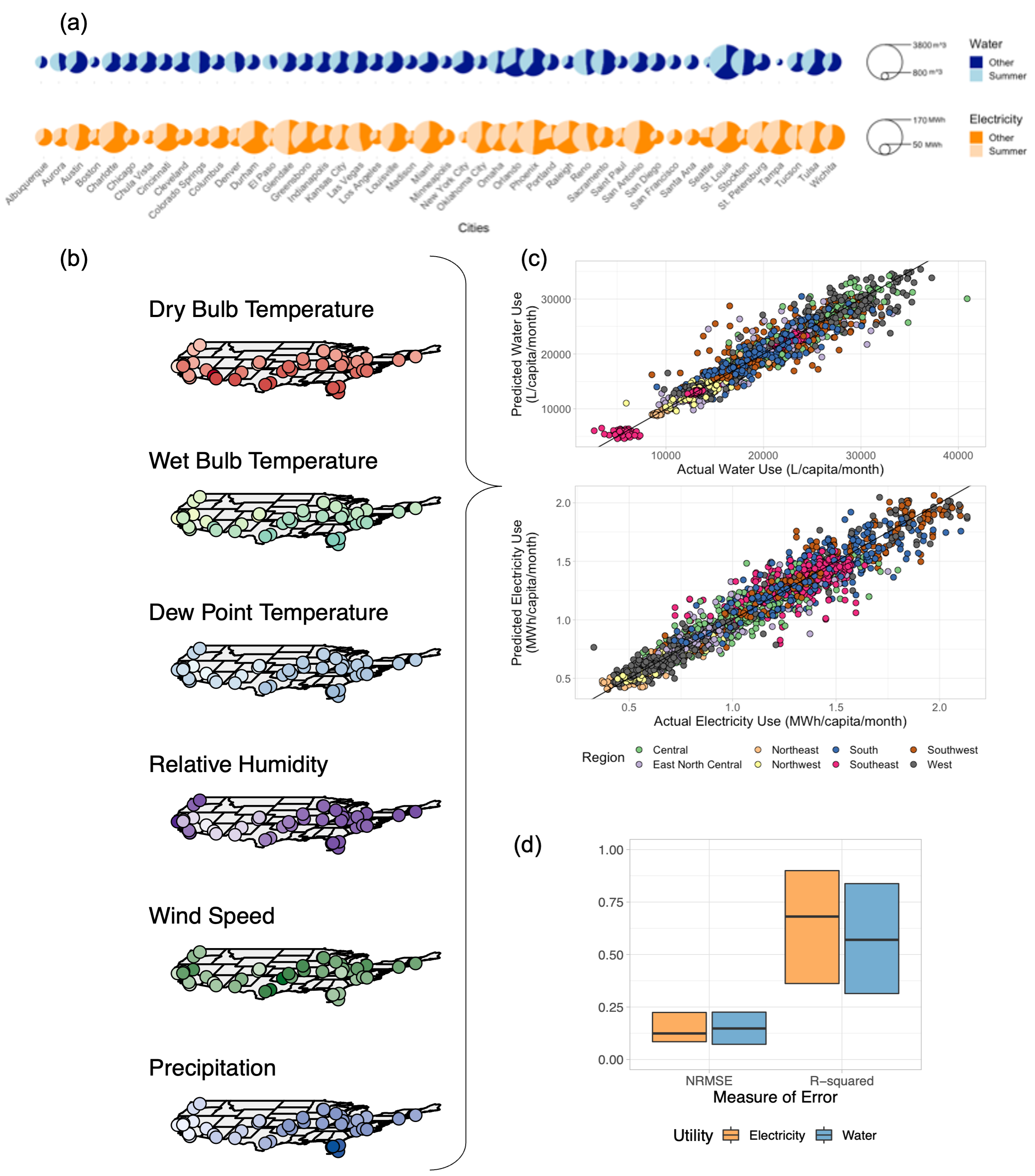}
\caption{(a) Total water ($m^3$) and electricity (MWh) consumption compared to the summer-only consumption for each city. (b) A selection of the initial input variables. (c) The predicted climate-sensitive portion of the water and electricity use plotted against the actual data, with a 45\textdegree \hspace{2pt} line for reference. (d) Measure of error, plotted as a box plot to represent the variance across the 46 cities. The lower bound represents the minimum value, the upper bound represents the maximum value, and the crossbar represents the median.}
\label{fig1}
\end{figure}

\subsection*{Methods}
The modeling framework was adapted from \citet{obringer2020a}, who projected the water and electricity demand into the future for six cities in the Midwestern United States. The framework is centered around a statistical learning algorithm, known as multivariate tree boosting \citep{miller2016}, which is capable of simultaneously predicting multiple variables. See the Supplementary Information for an algorithmic representation of this method. The algorithm also accounts for the interdependency between the response variables to improve the predictive accuracy, making the water-electricity nexus an ideal study application for the model. This work expanded the study by \citet{obringer2020a} to the entire continental United States, as well as changed several procedures within the framework. Namely, the model was first developed at a regional scale, grouping the cities based on the NOAA Climate Zones \citep{karl1984}. Within this step, we performed variable selection to reduce the number of variables within each region. Given that climatic patterns often evolve at the region-level, this step was important for determining which variables were important throughout a region, before delving into the city-level analysis. This regional analysis resulted in 4-6 important variables per region, rather than the original 17---reducing the complexity and improving model accuracy. See Supplementary Figures S1 and S5 for the important variables selected for each region.

Following the region-level variable selection, we performed individual city analyses with the previously selected variables. First, to avoid overfitting, the model was trained and rigourously tested via conducting randomized 5-fold cross-validation using observational climate data collected from NARR and water and electricity consumption data from local utilities. Then, using this trained model, we substituted the observed climate data with NARR climate data from the analogs, creating projections of water and electricity demand. The projections were conducted for two scenarios: RCP8.5 (high warming) and RCP4.5 (moderate warming), based on the ensemble analogs presented by \citep{fitzpatrick2019}. Finally, we evaluated the percent change between the analog-based projection and observed model results for each city. For more information on the methods used in this study, see the Supplementary Information, as well as Figure S4. 

These projections represent the most likely climate (based on the ensemble of 27 GCMs) for a given city in the year 2080 \citep{fitzpatrick2019}. The results represent the change to the per capita water and electricity demand in the selected cities, if everything else (e.g., technology, socioeconomic status, demographics, etc.) should stay the same. Although this is a large assumption, understanding the climate-only impacts can provide information on how much society needs to do to offset the changes. In other words, any increase in demand caused by climate needs to be met by the supply, or offset by increases to efficiency or conservation behaviors. This knowledge is beneficial to utility managers and policymakers that are implementing equitable policies to reduce climate impacts to infrastructure and seek to protect the most vulnerable groups. Moreover, the work is relevant for climate communication, as the use of climate analogs is easily understood by researchers, practitioners, and citizens alike. Finally, although we focused on 46 US cities, this model can be applied to a number of cities and regions, provided the data is available. The framework itself can be scaled up to regions or transferred to other cities, states, or countries.

\section*{Results}
The methodology of this work was composed to two modeling steps: (1) using observational data for model training, and (2) using climate analog data to investigate the impact of future climate change. Throughout the remainder of this paper, we will refer to the first step as a `predictive model' and the second step as a `projection model' to differentiate between the two steps. The predictive model was built using observational data from the North American Regional Reanalysis (see Methods and Figure \ref{fig1}b for input variables). In particular, we considered the summer months (June - September), which account for 30-50\% of the total annual water and electricity demand (Figure \ref{fig1}a). 

To measure model performance of the prediction model, we calculated $R^2$ and normalized root mean squared error (NRMSE). In terms of $R^2$, the median value for climate-driven portion of electricity (water) consumption was 0.70 (0.56), suggesting that the model does explain a notable percentage of variance for the majority of the cities. However, the values did range from 0.36 to 0.90 (0.31 to 0.84) for electricity (water) consumption indicating the varying degrees of climate sensitivity of demand across the US. In other words, it indicates that there are some cities that are not completely explained by the climate variables. Figure~\ref{fig1}c shows that in terms of water consumption, some cities in the Southeast, as well as the Southwest do not have as close of a fit as other areas. However, similar work has shown lower values of $R^2$ \citep{obringer2019c, obringer2020a}, suggesting that it is not uncommon to see low $R^2$ values when modeling the climate-sensitive portion of the water-electricity nexus. Nonetheless, $R^2$ is not a measure of error and is therefore rarely used to evaluate \textit{predictive} performance, which is usually characterized through measures such as RMSE or MAE. Here, we consider NRMSE as the primary measure of predictive error. 

Overall, the predictive model reliably characterized the historical records of the climate-sensitive portion of the coupled water and electricity demand, with a median NRMSE of 0.12 (0.15) for electricity (water) demand (see Figure~\ref{fig1}d). This suggests that the median error for both water and electricity consumption was below 15\%. In general, there was less variation in the NRMSE than the $R^2$, indicating that the \textit{predictive} performance of the model is relatively stable throughout the cities studied. This further indicates the transferability of the model, particularly if the goal is to predict the climate-sensitive portion of the water and electricity consumption. Going forward, this model was used to project the coupled water and electricity demand under the high emissions climate change scenario. 

The demand projections were made by leveraging climate analogs as a source of future climate data. The analogs were initially determined by Fitzpatrick and Dunn (2019) by calculating the sigma dissimaliarity for temperature (minimum and maximum) and precipitation across 540 North American cities \citep{fitzpatrick2019}. The authors used an ensemble of 27 GCMs to determine the analog for each city under RCP4.5 and and RCP8.5 \citep{fitzpatrick2019}. Generally, the authors aimed to find the location with less than $2\sigma$ difference between the cities, although some cities did have differences up to $8\sigma$, depending on the GCM \citep{fitzpatrick2019}. For example, New York City's climate analog for the year 2080 (RCP8.5) is Jonesboro, Arkansas \citep{fitzpatrick2019}. This means that in 2080, New York City can expect a climate similar to present-day Jonesboro, if there is no global reduction in emissions. In order to obtain future climate data for New York City, therefore, we collected observed climate data from Jonesboro (2007-2018) using the NARR dataset. In this way, we circumvent the need for downscaled GCM data, while making reasonable future projections of water and electricity demand. The results of the projections are shown in Figure~\ref{fig2}b and d. See Supplementary Figure S2 for the projections under a moderate emissions scenario (RCP 4.5).

\begin{figure}[t!]
\centering
\includegraphics[width=\linewidth]{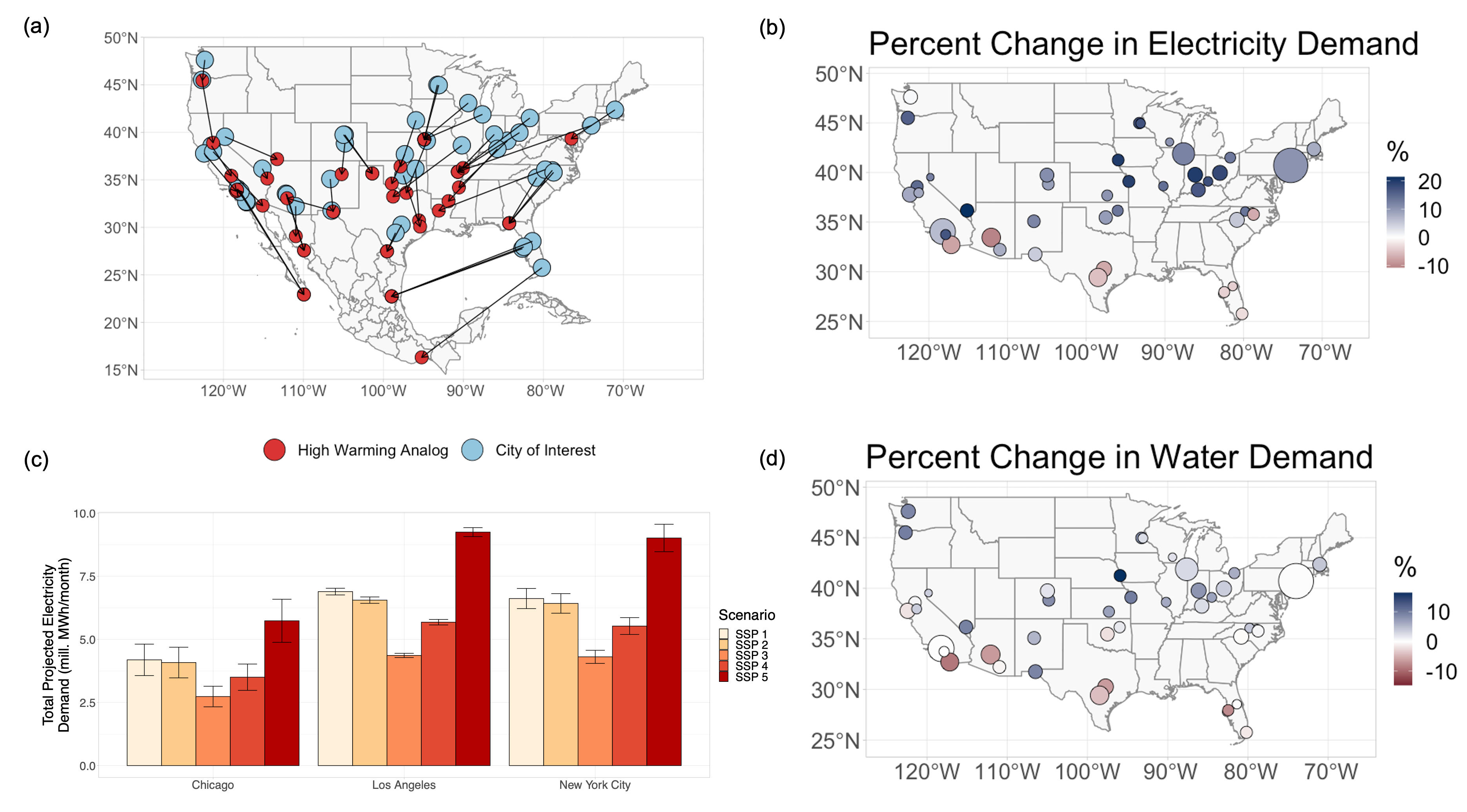}
\caption{(a) Target cities and their high warming analogs \citep{fitzpatrick2019}. (b) Projected changes to electricity demand across the 46 cities (sized by population). (c) Total projected electricity demand increase by 2080 in three major cities under the five Shared Socioeconomic Pathways (SSPs) \citep{hauer2019}. (d) Projected changes to the water demand across the 46 cities (sized by population).}
\label{fig2}
\end{figure}

\subsection*{Regional trends in the projected water and electricity
demand}

The majority of this increase is occurring in the Midwestern cities, particularly with respect to electricity use. Previous work has shown significant increases in electricity use, as well as air conditioning use in particular, within the Midwest region \citep{obringer2020a,obringer2022}. Our results align with these previous findings, indicating close to 20\% increase in the climate-sensitive portion of electricity demand by 2080, solely due to climate warming. Likewise, water consumption is expected to increase in many Midwestern cities. These projected increases are likely due to the analogs, which are generally located in the Southern US. This means that the average future under high warming for the Midwest will be warmer and drier, leading to increased demand for water and electricity within the residential sector. Previous work in the Midwest found similar results.

In particular, an analysis by \citet{obringer2020a} projected a 20\% ($\pm$15\%) increase in Chicago's electricity demand after surpassing the 2.0\textdegree C threshold, following the same water-energy nexus approach. The present analysis projected an increase of 12\% for the city of Chicago's electricity use, which is within the estimated range previously reported. The primary difference between the two methodologies was the use of climate analogs rather than downscaled GCM data. The fact that the present results for Chicago are within the range of GCM-based projections provides greater confidence in the reliability of this method, at least in Chicago. Nevertheless, additional work is needed to confirm the results for the rest of the cities, however, given that this method is significantly less computationally expensive and more user-friendly, it does signify the possibility of using climate analogs to discuss climate change impacts. 

In the Northeast, the electricity consumption is projected to increase under the high emissions scenario, while water is projected to remain relatively constant or increase slightly. This indicates that the average high emissions future has higher temperatures (which are often associated with higher electricity consumption), with limited changes to precipitation (which is often the primary climate driver of water consumption). In terms of water consumption, the mountain region and Western US are projected to see increases. Given the ongoing issues of drought within the Colorado River basin (as well as other Western US basins) \citep{guegan2012, madani2014, udall2017, fleck2021, gangopadhyay2022, lisonbee2022}, this increase could create further stress on the water system. Further, the increased consumption of electricity paired with this increase in water consumption could negatively impact hydropower in the Western US. Specifically, during periods of drought, the increased water consumption could create issues for the reservoirs, which may then limit the effectiveness of hydroelectric plant operations \citep{guegan2012, madani2014}. Understanding these interconnections are critical for building resilience and planning for future climate change.

Finally, there are some cities that are projected to see reductions in the climate-sensitive water and electricity demand, primarily in the Southern region, which tend to have analogs in more tropical regions (see Figure 1a). In general, the results indicate that under the high emissions scenario, the majority of the cities studied are projected to see increases in water and electricity demand up to 15\% and 20\% above current levels, respectively; although some cities are projected to see reductions in demand. These latter cases are found in the Southern regions of the country, demonstrating a north-south gradient to projected changes in water and electricity demand, which will be further discussed below.

\subsection*{Southern cities show decreases in climate-sensitive water-electricity demand}

Within the country as a whole, there is a gradient that separates the Southern cities from from the central and Northern cities. In particular, we see reductions in water and electricity demand below this gradient, with increases above it. This gradient is perhaps counterintuitive, as we would expect the warmer regions to get warmer (and potentially drier), which would lead to higher levels of electricity and water demand. However, by investigating the analogs of the Southern cities evaluated in this analysis, we find that the analogs are often more tropical, leading to increased precipitation and slightly cooler (average) dry bulb temperatures. Based on the relationship between climate and the water-electricity demand nexus, this leads the model to project reductions in the \textit{climate-sensitive} portion of water and electricity demand in these cities. It should be noted, however, that these results are based on the ensemble of 27 GCMs, thus there is some uncertainty in the initial climate analogs. The authors themselves discuss this uncertainty and point out that these counterintuitive results indicate that there are possible futures in which cities experience completely new climate zones \citep{fitzpatrick2019}. 

For example, our results indicate that the city of Phoenix (Arizona) may experience significant reductions in residential water and electricity demand---approximately 15\% in both sectors---due to the relationship between climate and the water-energy nexus, without accounting for changes in population, socioeconomic status, cultural norms, or other anthropogenic impacts, such as urban heat islands. This is a case in which the ensemble climate analog (Esperanza, Mexico) has a different climate than the city of interest. Given the growing population in Phoenix, as well as the ongoing mega-drought and periods of extreme heat, it represents an opportunity to evaluate the impact of model uncertainty on our results. 

In particular, Esperanza, Mexico is both cooler and wetter in the summer when compared to Phoenix. Thus, in our model, which leverages the relationships between climate variables and water-electricity demand, the occurrence of lower temperatures and more precipitation leads to lower consumption of both water and electricity. In particular, our analysis found that the most important variable for predicting the climate-sensitive portion of the water-energy nexus in Phoenix was the average dry bulb temperature (see Supplementary Figures S1 and S5 for important variables by city and region, respectively), which was nearly 20\% lower in Esperanza than Phoenix during the observation period. Given that higher average dry bulb temperatures tend to lead to higher electricity consumption within the observed values for Southwestern region (see Supplementary Figure S3), it is likely that this reduction in temperature leads to a projection of reduced electricity demand due to the climatic conditions. Similarly, an increase in precipitation would lead to a reduction in water consumption due to a reduction in the need for outdoor landscaping. 

While the model itself is built on these relationships, it is important to note that the uncertainty in the future climate analogs could lead to different values. For example, \citet{fitzpatrick2019} presented the results from 27 individual GCM-based analyses plus the ensemble. If we investigate the individual analogs for Phoenix, they range from Buckeye, Arizona (about 40 miles southwest of Phoenix) to several locations in the Sonora state of Mexico (including the cities of Nacozari de Garc\'{i}a, Hermosillo, and Esperanza). While the ensemble analog resulted in a projected 15\% decrease in water and electricity consumption, the individual analogs lead to different results. In particular, if we consider Buckeye, Arizona as the analog for Phoenix, the result is a 4.5\% (1.8\%) reduction in water (electricity) consumption. Likewise, if we evaluate the other Mexican cities, the result is a 6.3\% (10.0\%) reduction and a 3.5\% (12.5\%) reduction in water (electricity) consumption for Hermasillo and Nacozari de Garc\'{i}a, respectively. Thus, we can see that the different analogs lead to different magnitudes of change for the city of Phoenix, depending on the GCM used to derive the analog. This is representative of the uncertainty present within studies that leverage multiple GCMs. 

Likewise, the cities in Florida are projected to see reductions in our modeling results. In particular, the city of Tampa is projected to see close to a 10\% decrease in water demand, with only a slight decrease in electricity. The climate analog of Tampa is Ciudad Mante, Mexico \citep{fitzpatrick2019}, which receives more precipitation in the summer than Tampa. This increase in precipitation would likely lead to less outdoor water use, the main source of urban water demand during the summer months. Looking that the possible analogs for Tampa, they all fall along the Gulf coast in Mexico, ranging from Ciudad Mante (the ensemble analog) to Palenque in the state of Chiapas. Comparing Ciudad Mante to Palenque, both are tropical climates, however, the K\"{o}ppen climate designation for Ciudad Mante is tropical savanna while that of Palenque is tropical monsoon. These differences in climate are likely to cause differences in the projection of the climate-sensitive water and electricity demand in the city of Tampa. In fact, if we consider Palenque as the analog for Tampa, the resulting change is a 6.5\% (3.1\%) reduction in water (electricity) consumption. Although we did not investigate all 1,242 analogs (27 GCMs $\times$ 46 cities), it is likely that there would be similar results across the US, suggesting the importance of quantifying uncertainty in future climate analogs analyses.

Ultimately, the projections of water and electricity consumption that we presented here are based on the observed relationship between the climatic conditions and the water-electricity demand nexus in a given city, which may change in the future due to a number of factors (e.g., technology, cultural norms, or demographics). In this sense, it is important to acknowledge the uncertainty associated with the GCM-derived analogs, which were used to obtain these projections. The uncertainty associated with the 27 possible analogs could lead to some differences in projections, as discussed for the cities of Phoenix and Tampa. As the GCMs continue to evolve and improve, it is likely that the range of possible climate analogs will be reduced, leading to more certain results. Additionally, this approach does not account for extreme events, since we are focusing on projecting the monthly demand for an average month in future summers, nor does it assume any changes to the current demographics or cultures of the cities. One way to explore these additional impacts is through the Shared Socioeconomic Pathways (SSPs), which provide scenarios of future population growth.

\subsection*{Exploring the impacts of shared
socioeconomic pathways}
As shown in Figure~\ref{fig2}b, the majority of the cities included in the analysis are projected to see increases in electricity demand, including the most populous cities in the country: New York City (New York), Chicago (Illinois), and Los Angeles (California). These cities show little to no change in the water demand, likely due to their dense nature, which does not allow for much outdoor use, but with large increases in electricity use (6-12\%) per capita. To understanding the impact of population growth, we can evaluate the increase in electricity use based on the SSPs of the major cities mentioned above.

In terms of the total change in electricity demand, a 6-12\% increase in per capita electricity demand in these three major cities represents an increase of 2.5-9 million MWh in a given summer month, depending on the city and population growth scenario (see Figure~\ref{fig2}c). In 
Figure~\ref{fig2}c, we show the total monthly electricity increases in the counties where Chicago, Los Angeles, and New York City reside, based on the five SSPs \citep{hauer2019}. The SSP data is aggregated by county, not city, so the total electricity values are for the counties as a whole (Cook County for Chicago, the combination of Bronx County, King County, New York County, Queens County, and Richmond County for New York City, and Los Angeles County for Los Angeles). We make the assumption that the residents beyond the city limits will have the same increase in electricity demand, given the similarity in climatic changes that will be experienced. Within these pathways, SSP 1 represents sustainable development, while SSP 5 represents fossil-fueled development \citep{hauer2019}. As SSP 5 is more in line with the worst-case warming scenario, we can use the total electricity from that scenario to better understand the possible changes to the electricity demand patterns under extreme climate change. In particular, the results presented in Figure~\ref{fig2}c show that in 2080, Chicago will require 5.7 million MWh in a given summer month compared to the 4.3 million MWh generated now. Likewise, Los Angeles will require 9.2 million MWh (compared to 5.4 million MWh today) and New York City will need to provide 9.0 million MWh, compared to the 4.2 million MWh required currently. 

These increases represent a significant portion of the existing electricity generation capacity. For example, Chicago will have to provide an additional 1.4 million MWh in a given summer month, this the equivalent power output of the Hoover Dam over 127 days \citep{ritchie2017}. In terms of carbon emissions, this increase in electricity demand would lead to an additional 605,000 metric tons of CO$_2$e month, based on the 2022 national electricity mix \citep{usepa2020}. Similarly, following an increase in Los Angeles (county) electricity demand of 3.8 million MWh, nearly 1.6 million metric tons of CO$_2$e would be emitted \citep{usepa2020}. Finally, New York City is projected to experience an increase of 4.8 million MWh in electricity demand by 2080, this would emit 2.1 million metric tons of CO$_2$e in a given summer month \citep{usepa2020}. Together, these three cities would emit an additional 4.3 million metric tons of CO$_2$e a month, assuming there are no mitigation efforts. To sequester these additional emissions, the U.S. would need add 20,717 km$^2$ of forested land \citep{usepa2020}, enough to nearly cover the state of New Jersey. It is important to note that these comparisons assume no mitigation carbon emissions through the electricity mix, which is an unlikely scenario, particularly given these three cities' commitment to renewable energy \citep{johnson2008, garcetti2019, deblasio2019}. Additionally, as the SSPs represent different pathways for population growth, it is possible that society will follow a more sustainable path, limiting the total emissions, as shown in Figure 2c. Nonetheless, there is still value to assessing the worst-case scenario, as we will discuss below. 

\section*{Discussion}
Within this manuscript, we have primarily focused on a worst-case, high emissions scenario. This means that we not only used the RCP8.5 scenario for future global emissions, but also did not account for technological advances or adaptation (e.g., to reduce consumption). The primary aim of this study was to focus on the impact of climate change on water-electricity demand, so this worst-case scenario provides an opportunity to (1) evaluate the impacts solely attributable to climate change and (2) understand the magnitude of change that will be needed to avoid those impacts. Additionally, while the idea that there will be no technological advancements in the next 60 years is not plausible, the chance of a high emissions scenario remains possible \citep{schwalm2020}. In other words, if there is not global action to avert the climate crisis, then it is unlikely to matter what cities or even states, like California, do in terms of mitigation. Reducing emissions is a global problem, so even if localities implement climate adaptation and mitigation plans, without large-scale action around the world, emissions will continue to rise and the climate crisis will worsen. As such, it is critical that cities account for potential climate-induced shifts in water and electricity demand when developing future plans in addition to working towards mitigation. In particular, the unaccounted for change in demand could make it more difficult for cities to reach renewable energy targets. For example, Los Angeles plans to achieve 100\% renewable energy by 2050 \citep{garcetti2019}. An increase of 3.8 million MWh across the county would require 14,000 additional 1.5 MW wind turbines, approximately 20\% of the current operational stock in the United States, just for the relatively small area servicing Los Angeles County. If cities are not prepared for these changes, they may fall back on carbon-intensive electricity generation, which will ultimately exacerbate the climate crisis. 

While a detailed overview of the injustice implications of our findings is beyond the scope of this study, it is important to note that the impacts of unmitigated climate change to the water and electricity grid are likely to have a greater effect on vulnerable groups, exacerbating existing environmental injustices \citep{bednar2020,obringer2022}. The results presented here represent the average increase over the course of a single summer month. There will be, however, extreme events, such as droughts or heatwaves, that lead to peaks in water and electricity consumption beyond our analysis. These events put a strain the infrastructure systems, with managers and operators often resorting to conservation mandates \citep{mehran2017,wendt2021} or, in the case of the electric grid, rolling blackouts \citep{vanvliet2012,yalew2020}. Previous research has shown that vulnerable groups, such as lower income families, the elderly, or racially/ethnically marginalized people, experience more extreme effects during heatwaves \citep{khosla2020}. This increased vulnerability to heat-related events can be attributed to a number of factors, from the urban heat island effect to the inability to afford rising bills \citep{drehobl2016, berry2018, sanchez-guevara2019, graff2020, bednar2020}. Should utilities and management agencies fail to account for the climate-induced shifts in demand, as the results presented here suggest are likely, the resulting supply deficits should result in increased vulnerability to heat-related issues, as well as other impacts. In this sense, it is critical that society work to not only mitigate the climate crisis, but also to prepare our infrastructure for the worst case scenario to ensure equitable access to water and electricity for future generations. To do this effectively, we need to continue to expand our modeling capabilities, as well as focus on finding user-friendly and computationally inexpensive ways to model climate impacts on infrastructure systems. 

\subsection*{Study limitations and areas for future work}
In this analysis, we focused on the summer months, showing a north-south gradient in terms of climate-induced shifts in water and electricity consumption. Given that climate change is likely to lead to warmer winters, there is a need to further understand how the climate-sensitive portion of water and electricity consumption will change during this season as well. In fact, in many locations, the higher temperatures might lead to an increased need for electricity-based space-cooling, where previously, there was a greater need for natural gas-based space-heating. The difference between summer and winter months may be particularly evident below the gradient, in the Southern states. For example, while Esperanza, Mexico is cooler and wetter than Phoenix in the summer, it is warmer and drier in the winter. This could lead to an unbalanced demand structure, which may further stress infrastructure systems. Likewise, Tampa could see wetter summers and drier winters, given their ensemble analog of Ciudad Mante, Mexico. This indicates a need to further evaluate the climate change impacts during winter on the climate-sensitive portion of the water-electricity nexus, particularly in the Southern cities considered in
this study.

Additionally, the climate analogs were initially developed using the CMIP5 suite of GCMs \citep{fitzpatrick2019}. As the CMIP6 models become more commonplace and downscaled, bias-corrected data becomes more readily available from these next generation GCMs, it might be beneficial to develop new climate analogs. Additionally, in the original study, the authors only considered maximum temperature, minimum temperature, and precipitation to determine the most likely analog. However, recent work has indicated that relative humidity and other measures of temperature play a major role in climate change impacts \citep{maia-silva2020, kumar2020, obringer2022} therefore, including these climatic variables might improve the climate analogs development, particularly for use in impact assessments. While the aim of this study was not to criticize or improve upon the climate analogs, making these improvements would like help to build up the use of climate analogs as both a modeling and communication tool, as well as provide greater certainty around future projections in the climate change impact space.

\section*{Conclusion}
In this study, we build a scalable and transferable model for projecting the coupled water and electricity demand into the future under climate change. In particular, we leverage climate analogs as a modeling tool to make future projections across 46 cities in the United States under a high emissions climate change scenario. Moreover, we use a water-energy nexus framing, which models the two sectors simultaneously and has been shown to be more representative of the real world systems \citep{bartos2014c,obringer2019c}. The results indicate increases in climate-sensitive water and electricity demand up to 15\% and 20\%, respectively. In the absence of proactive mitigation strategies, the unanticipated increases in demand could lead to supply shortages and complicate the transition to clean energy. Moreover, previous research has indicated that vulnerable groups, such as low income families or the elderly, are likely to experience more intense impacts in the event of supply shortages. As such, preparing for the climate-induced shifts in water and electricity demand is not only an issue of climate change adaptation, but also one of environmental justice.

Going forward, it is important to continue to develop future projections of water and electricity demand that not only account for the interdependencies between the two sectors, but also the impact that climate change will have on the coupled demand structure. Ultimately, this coupled demand will feedback into the supply side, which is also deeply interconnected (e.g., water is necessary for many modes of electricity generation, so higher demand for electricity not only leads to more water demand at the end-use, but also at the generator). In this sense, by making these interconnected projections, we can also encourage interconnected action, such as the electricity utilities working to encourage water conservation as a way to maintain adequate reservoir levels for hydropower generation. This is also critical for the shift to renewable energy, since planning electricity generation requires accurate representations of the demand structure. For example, energy system plans that do not account for the climate impact on demand or that this impact is likely to effect the water sector, which in turn will effect the electricity sector, may not achieve the necessary capacity to provide adequate supply for future generations. In this sense, the modeling of climate impacts through a water-energy nexus framing is critical to furthering climate change mitigation and adaptation efforts. Moreover, the use of climate analogs as a proxy for future climate data can greatly reduce the computational needs, potentially providing practitioners with a more interpretable way to model and communicate the impacts of climate change on their local systems. Finally, although this work focused on the climate-sensitive portion of water and electricity demand, the methods can be easily integrated with work investigating the impact of technological change and population growth, through SSPs, for example, as was demonstrated for three major cities above. Additionally, the model can be scaled to encompass regions or transferred to other cities, states, or countries, provided the data is available. This type of interdisciplinary work is critical for building resilience to climate change and finding equitable solutions that are accessible to local governments and organizations.

\bibliographystyle{plainnat}
\bibliography{bibfile} 

\end{document}